\documentclass[aps,prb,superscriptaddress,twocolumn,amsmath,amssymb,amsfonts,10pt,nofootinbib]{revtex4-2}

\usepackage[T1]{fontenc}
\usepackage{graphicx}
\usepackage{enumitem}
\usepackage{amsmath}
\usepackage{amsfonts}
\usepackage{amssymb}
\usepackage{mwe}
\usepackage{color}
\usepackage[dvipsnames]{xcolor}

\usepackage{simplewick}
\usepackage[colorlinks,bookmarks=false,urlcolor=blue,citecolor=blue,linkcolor=blue]{hyperref}
\usepackage[capitalise]{cleveref}
\usepackage{comment}
\newcommand{\setsmalltitle}[1]{{\textit{#1}.}}

\newcommand{\ee}{\mathrm{e}} %For exponentials
\newcommand{\ket}[1]{\left| #1 \right\rangle}

\usepackage{tikz}

\DeclareRobustCommand{\circleletter}[1]{%
  \raisebox{-1.5pt}{\tikz[baseline={(char.south)}]{%
    \node[shape=circle,draw,inner sep=0.5pt] (char) {#1};%
  }%
  }%
}

\begin{document}

\title{Mass-gap functional determinant approach for mobile Fermi polarons}

\author{Emilio Ramos Rodr\'iguez}
\thanks{These authors contributed equally to this work.}
\affiliation{Université Paris Cité, Laboratoire Matériaux et Phénomènes Quantiques (MPQ), CNRS, F-75013, Paris, France}

\author{Eugen Dizer}
\thanks{These authors contributed equally to this work.}
\affiliation{Institut für Theoretische Physik, Universität Heidelberg, 69120 Heidelberg, Germany}

\author{Xin Chen}
\affiliation{Institut für Theoretische Physik, Universität Heidelberg, 69120 Heidelberg, Germany}

\author{Richard Schmidt}
\affiliation{Institut für Theoretische Physik, Universität Heidelberg, 69120 Heidelberg, Germany}
\affiliation{Center for the Physical Foundations of Computation, Heidelberg University, Heidelberg, Germany}

\begin{abstract}
We extend the functional determinant approach (FDA), previously restricted to static impurities, to the case of finite-mass impurities by using the mass-gap description of Fermi polarons [Phys.~Rev.~Lett.~\textbf{135}, 193401 (2025)]. The quadratic structure of the mass-gap model enables the exact evaluation of many-body spectra and Ramsey dynamics for mobile impurities. We show that this \textit{mass-gap FDA} smoothly interpolates between the Fermi-edge singularity for infinitely heavy impurities and the emergence of quasiparticle weight for finite impurity mass. Our results demonstrate that the mass-gap FDA provides a transparent and computationally efficient framework to describe the quantum dynamics of mobile impurities.
\end{abstract}

\maketitle

%%%%%%%%%%%%%%%%%%%%%%%%%%%%
%\setsmalltitle{Introduction}
Fermi polarons --- mobile impurities dressed by 
particle-hole excitations of a surrounding Fermi sea 
--- arise across a wide range of quantum many-body 
systems, from ultracold atomic 
mixtures~\cite{Schirotzek:2009exp,Kohstall2012,
Massignan:2014,Cetina2016,Scazza2017,Yan2019,
Ness:2020PolMol,Baroni2024,Baroni2024FP,
massignan2025polarons} to charge carriers in 
two-dimensional 
semiconductors~\cite{Sidler:2017,Efimkin:2018XPol,
Goldstein2020,Efimkin2021,imamoglu2021exciton,Rana2021,Xiaoqin2023}. The interplay between 
impurity motion and interaction with the surrounding Fermi sea gives 
rise to rich quasiparticle physics, which is controlled 
by the impurity mass.

In the limit of an infinitely heavy impurity, the 
problem reduces to a quadratic theory for the fermions 
that admits an exact solution via the functional 
determinant approach 
(FDA)~\cite{Levitov1993,Klich2003,Abanin2005,Knap:2012OC,Schmidt2018,Wang:2022,
Gievers2024Rydberg}. Using the FDA, one gains access to the 
full quantum many-body dynamics and impurity spectral 
function, capturing both the ground-state and 
excitation branches of the spectrum with the 
Fermi-edge singularity characteristic of the Anderson 
orthogonality catastrophe 
(OC)~\cite{Anderson:1967,Nozieres:1969}. The 
static-impurity FDA serves as an exact benchmark for 
variational 
methods~\cite{Chevy2006,Combescot2007,Combescot2008,
Punk:2009PolMol,Mora:2009,Schmidt2012,
Ngampruetikorn_2012,Parish:2016,Kain:2017HF,Liu:2019ThermoVA,Adlong:2020,Dolgirev:2021,Qu:2026} 
and diagrammatic 
techniques~\cite{Prokofev:2008,Prokofev:2008_2,
Schmidt:2011,Vlietinck2014diag,Kroiss2015,Hu2024}. 
However, its application to mobile impurities of 
finite mass has remained elusive: the Lee-Low-Pines 
transformation~\cite{Lee:1953}, which decouples the 
impurity from the bath, generates a quartic interaction 
term in the fermionic Hamiltonian and prevents a direct 
application of the FDA. This leaves the connection 
between the spectral properties of the OC and the quasiparticle picture of Fermi polarons as an outstanding theoretical 
challenge~\cite{Pimenov2017,Pimenov:2018,Gievers2025}.

Recently, the mass-gap description of mobile impurities 
in Fermi gases was introduced~\cite{Chen:2025}, which 
provides a new model to address this difficulty by means of an operator 
reordering of the quartic Lee-Low-Pines terms. The 
resulting quadratic part of the Hamiltonian encodes the 
impurity recoil in a modified fermionic dispersion 
relation featuring an energy gap $\Delta(M)$ at the 
Fermi surface --- the mass gap. This formulation opens 
the door to describe the quantum dynamics of finite-mass impurities with the FDA.

In this Letter, we show that the Hamiltonian of the 
mass-gap model enables a natural extension of the FDA 
to mobile impurities, which we term the 
\textit{mass-gap FDA}. Using the mass-gap FDA, we 
compute absorption spectra and real-time Ramsey dynamics that smoothly connect 
the static impurity limit to the mobile impurity regime 
of well-defined Fermi polaron quasiparticles. The mass 
gap $\Delta(M)$ acts as an infrared regulator for the 
OC, introducing a timescale governing the formation of 
quasiparticles. Furthermore, we identify the 
microscopic origin of spectral features including the 
dark continuum~\cite{Goulko2016} and the molecule-hole scattering 
continuum~\cite{Bruun:2010PolMol,Parish:2011_2DPolMol,Cui2020,Cui:2021PolMol,Parish2021}. We compare our 
results to the variational Chevy ansatz~\cite{Chevy2006} and 
the conventional static-impurity FDA for the experimentally relevant 
Li-Cs mass ratio~\cite{Rautenberg:2026}.
\begin{figure*}
    \centering 
    \includegraphics[width=\linewidth]{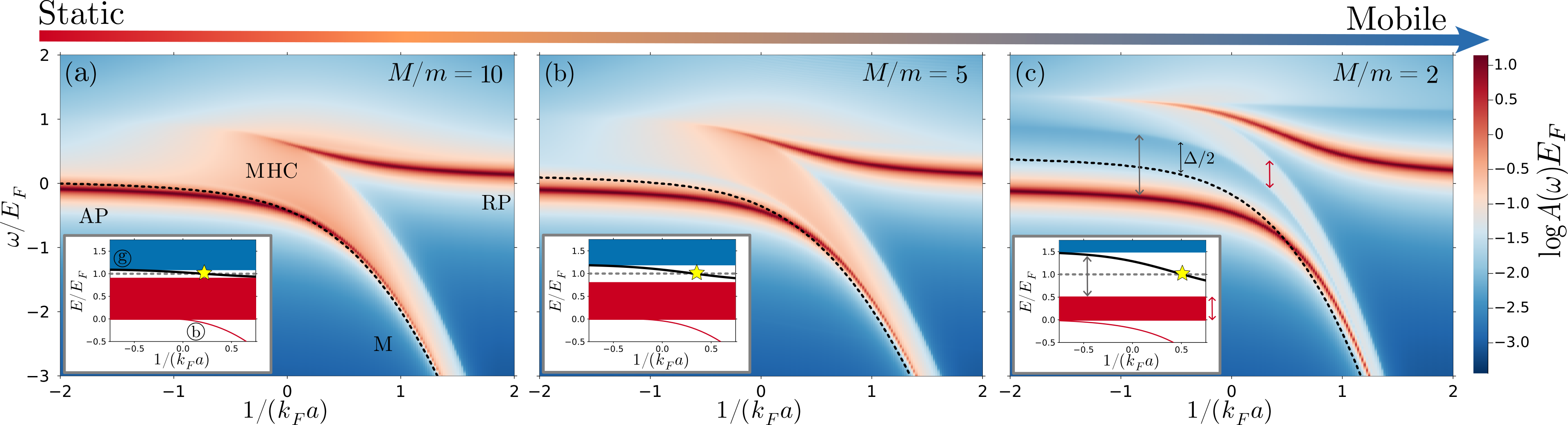}
    \caption{\textbf{Single- and many-body spectrum of the mass-gap model.}
    Many-body absorption spectrum $A(\omega)$ of the Fermi 
    polaron at zero momentum obtained from the mass-gap FDA
    for different impurity-fermion mass ratios (a) $M/m=10$, (b) $M/m=5$, and (c) $M/m=2$. The spectra show three distinct features: the attractive polaron (AP) and repulsive polaron (RP) branches, and the molecule-hole continuum (MHC). The dashed black line denotes the energy of the molecule (M) obtained from the mass-gap model~\cite{Chen:2025}. Inset: Single-particle energy spectrum of the 
    mass-gap Hamiltonian $\hat{\mathcal{H}}$ obtained from 
    exact diagonalization, as a 
    function of interaction strength $1/(k_F a)$. The discrete 
    bound state $\circleletter{b}$ and in-gap state $\circleletter{g}$ are highlighted, 
    together with the upper and lower scattering continua 
    separated by the mass gap $\Delta(M)$ at the Fermi energy 
    $E_F$. The polaron-to-molecule transition occurs when the 
    in-gap state crosses $E_F$ (marked by a yellow star).}
    \label{fig:1}
\end{figure*}
%

%%%%%%%%%%%%%%%%%%%%%%%%%%%%
\setsmalltitle{Mass-gap model}
We consider a single impurity of mass $M$ immersed in 
a three-dimensional Fermi gas of particles with mass 
$m$ at zero total momentum, interacting via short-range 
contact interactions. In the Lee-Low-Pines (LLP) 
frame~\cite{Lee:1953}, the impurity is decoupled from the bath and the system is described by the Hamiltonian~\cite{Kain:2017HF},
\begin{align} \label{eq:LLP-Hamiltonian}
    \hat{H}_{\mathrm{LLP}} =& \,\sum_{\mathbf{k}} \left(\frac{\mathbf{k}^2}{2m}+\frac{\mathbf{k}^2}{2M}\right)\hat{c}^{\dagger}_{\mathbf{k}}\hat{c}_{\mathbf{k}} + \frac{g}{\mathcal{V}} \sum_{\mathbf{k},\mathbf{k}'} \hat{c}^{\dagger}_{\mathbf{k}}\hat{c}_{\mathbf{k}'} \notag \\
    &+ \frac{1}{2M} \sum_{\mathbf{k},\mathbf{k}'} \left(\mathbf{k}\cdot\mathbf{k}'\right) \hat{c}^{\dagger}_{\mathbf{k}}\hat{c}^{\dagger}_{\mathbf{k}'}\hat{c}_{\mathbf{k}'}\hat{c}_{\mathbf{k}} \,.
\end{align}
Here $\hat{c}^{\dagger}_{\mathbf{k}}$ 
($\hat{c}_{\mathbf{k}}$) creates (annihilates) a bath 
fermion of momentum $\mathbf{k}$, $g$ is the bare coupling 
constant related to the $s$-wave scattering 
length $a$ via $1/g = m_r/(2\pi a) - m_r\Lambda/\pi^2$ 
with reduced mass $m_r=mM/(m+M)$ and momentum cutoff $\Lambda$, and $\mathcal{V}$ is the quantization volume. While the impurity has been decoupled, the resulting quartic term in the fermionic operators prevents a direct application of the FDA, which requires a quadratic Hamiltonian.

By normal ordering the quartic term with respect to the filled Fermi sea, the Hamiltonian~\eqref{eq:LLP-Hamiltonian} can be written as $\hat{H}_{\mathrm{LLP}} = \hat{\mathcal{H}} + \hat{\mathcal{H}}_{\mathrm{int}}$, where $\hat{\mathcal{H}}$ is quadratic and $\hat{\mathcal{H}}_{\mathrm{int}}$ contains the residual reordered quartic terms. 
The quadratic part is exactly the mass-gap Hamiltonian introduced in~\cite{Chen:2025},
\begin{align} \label{eq:quadratic-Hamiltonian}
    \hat{\cal H} =  \sum_{\mathbf{k}} E_{\mathbf{k}} \hat{c}^{\dagger}_{\mathbf{k}}\hat{c}_{\mathbf{k}} + \frac{g}{\mathcal{V}} \sum_{\mathbf{k},\mathbf{k}'} \hat{c}^{\dagger}_{\mathbf{k}}\hat{c}_{\mathbf{k}'} \,,
\end{align}
with the modified fermionic dispersion relation,
\begin{align} \label{eq:modified-dispersion}
    E_{\mathbf{k}} =
    \begin{cases}
        \frac{\mathbf{k}^2}{2m} - \frac{\mathbf{k}^2}{2M} & \quad\mathrm{for}\quad |\mathbf{k}|<k_F \,, \\[1ex]
        \frac{\mathbf{k}^2}{2m} + \frac{\mathbf{k}^2}{2M} & \quad\mathrm{for}\quad |\mathbf{k}|>k_F \,.
    \end{cases}
\end{align}
The modified dispersion encodes the finite impurity mass through an energy gap $\Delta(M) = k_F^2/M$ at the Fermi momentum $k_F$. Throughout this work, we take the quadratic mass-gap Hamiltonian $\hat{\cal H}$ as the starting point and treat $\hat{\mathcal{H}}_{\mathrm{int}}$ as a residual interaction. The quadratic mass-gap Hamiltonian provides a controlled description for heavy impurities.
In particular, in the limit $M \to \infty$, the gap closes and the standard problem of a static impurity is recovered exactly.

As shown in Ref.~\cite{Chen:2025} and illustrated in the insets of
Fig.~\ref{fig:1}, the single-particle spectrum of 
$\hat{\mathcal{H}}$ features two discrete states in 
addition to the scattering continua: a bound state $\circleletter{b}$
at sufficiently strong attraction, 
and an in-gap state $\circleletter{g}$ inside the mass 
gap (black line in the insets), which exists for any finite impurity mass and 
interaction strength. The three configurations obtained 
by occupying these states correspond to the attractive 
polaron (AP, $\circleletter{b}$ occupied if present), the molecule (M, $\circleletter{g}$ and $\circleletter{b}$ occupied), and the repulsive polaron (RP, $\circleletter{b}$ unoccupied $\circleletter{g}$ occupied)~\cite{Chen:2025}. When the 
mass gap closes, for $M \to \infty$, the in-gap state disappears, and AP and M become identical. The mass gap thus 
acts as a regulator that continuously connects the 
quasiparticle regime for mobile impurities to the OC for static impurities.

\setsmalltitle{Mass-gap FDA}
The quadratic structure of the mass-gap Hamiltonian $\hat{\mathcal{H}}$ enables 
the application of the functional determinant 
approach (FDA)~\cite{Knap:2012OC,Schmidt2018,Wang:2022}. The impurity 
spectral function $A(\omega)=\frac{1}{\pi}\, \text{Re}\,
\int_0^{\infty} dt\, \ee^{i \omega t}\, S(t)$ is obtained 
from the Fourier transform of the Ramsey signal,
\begin{align} \label{eq:Ramsey-signal}
    S(t) = \langle \ee^{i\hat{\mathcal{H}}_0 t} 
    \ee^{-i\hat{\mathcal{H}} t} \rangle +\mathcal{O}(\hat{\mathcal{H}}_{\mathrm{int}}^2)\,,
\end{align}
where $\hat{\mathcal{H}}_0 = \sum_{\mathbf{k}} E_{\mathbf{k}} 
\hat{c}^{\dagger}_{\mathbf{k}} \hat{c}_{\mathbf{k}}$ is the 
non-interacting part of the mass-gap Hamiltonian and the 
expectation value is taken with respect to the non-interacting 
Fermi sea $|\mathrm{FS}\rangle$. As shown in the Supplemental Material (SM)~\cite{Supplemental}, corrections to the mass-gap model arise only at second order in perturbation theory with respect to $\hat{\mathcal{H}}_{\mathrm{int}}$, and are additionally suppressed by powers of $1/M$.

Since $\hat{\mathcal{H}}$ is 
quadratic, $S(t)$ can be decomposed using Wick's theorem and evaluated at any given order within the FDA framework. Neglecting the higher-order terms in Eq.~\eqref{eq:Ramsey-signal}, $S(t)$ retains the usual form~\cite{Knap:2012OC,Schmidt2018}
\begin{align} \label{eq:FDA}
    S(t) = \det\left[ 
    1 - n_F(\hat{h}_0) + n_F(\hat{h}_0)\, 
    \ee^{i\hat{h}_0 t} \ee^{-i\hat{h} t} 
    \right] \,,
\end{align}
where $n_F(\hat{h}_0)$ is the Fermi distribution, and $\hat{h}_0$ and $\hat{h}$ are the single-particle 
operators of $\hat{\mathcal{H}}_0$ and 
$\hat{\mathcal{H}}$, respectively. Although the FDA has 
previously been restricted to static 
impurities, the mass-gap model makes it 
applicable to mobile impurities by conserving the quadratic structure of the many-body problem, while still capturing the motion of the impurity.

%%%%%%%%%%%%%%%%%%%%%%%%%%%%
\setsmalltitle{Polaron spectra: from static to mobile impurities} In Fig.~\ref{fig:1}, we show the impurity absorption spectrum at zero temperature for different mass ratios $M/m$. Similar to the infinite-mass case~\cite{Knap:2012OC,Wang:2022}, the spectra obtained from the mass-gap model exhibit three prominent features: the attractive polaron (AP),  
repulsive polaron (RP), and a molecule-hole continuum (MHC) located between them. The energy scales associated with
these features are determined by the single-particle spectrum of $\hat{\mathcal{H}}$, and therefore explicitly depend on the 
impurity mass through the mass gap $\Delta(M)$.

For heavy impurities [see Fig.~\ref{fig:1}(a)], the spectrum closely resembles the infinite-mass limit, as the mass-gap is barely resolved. As the mass ratio decreases, i.e., as the impurity becomes lighter, the spectrum progressively evolves, revealing the effects of the mass-gap [see Figs.~\ref{fig:1}(b) and (c)]. In the MHC particularly, the mass-gap becomes most evident. As shown in Fig.~\ref{fig:1}(c), the emergent gap (gray arrow) between the AP and the onset of the lower edge of the MHC is given by the energy separation between the lower band of the single-body spectrum and the in-gap state $\circleletter{g}$. At the same time, the width of the MHC corresponds to the width of the lower band (red arrows). The onset energy of the continuum, $E_{\mathrm{onset}}^{\mathrm{MHC}}$, can then be related to the molecular energy $E_{\mathrm{mol}}$ as ${E_{\mathrm{onset}}^{\mathrm{MHC}}=E_{\mathrm{pol}}+E_{g}-(E_F-\Delta/2)=E_{\mathrm{mol}}+\Delta/2}$, where $E_g$ is the energy of the in-gap state~\cite{Chen:2025}. Consequently,
the energy splitting between the two features is exactly 
$\Delta(M)/2$ (black arrow), independent of the interaction strength. Remarkably, the precise measurement of the MHC and the AP thus provides a direct spectroscopic signature of the mass gap, as well as the molecular state.

Moreover, the microscopic origin of the MHC is 
captured within the mass-gap description from its single-body spectrum: since the in-gap state $\circleletter{g}$ is related to the molecular configuration at the single-particle level, particle-hole excitations to $\circleletter{g}$ describe a molecule dressed by a hole, giving rise to the name \textit{molecule-hole} continuum. In the limit $M \to \infty$, the in-gap state disappears and the molecule-hole continuum merges 
with the rest of the spectrum to form the power-law tail of the Fermi-edge singularity, consistent with the OC scenario~\cite{Knap:2012OC,Schmidt2018}.

\begin{figure}[t]
    \centering
    \includegraphics[width=\linewidth]{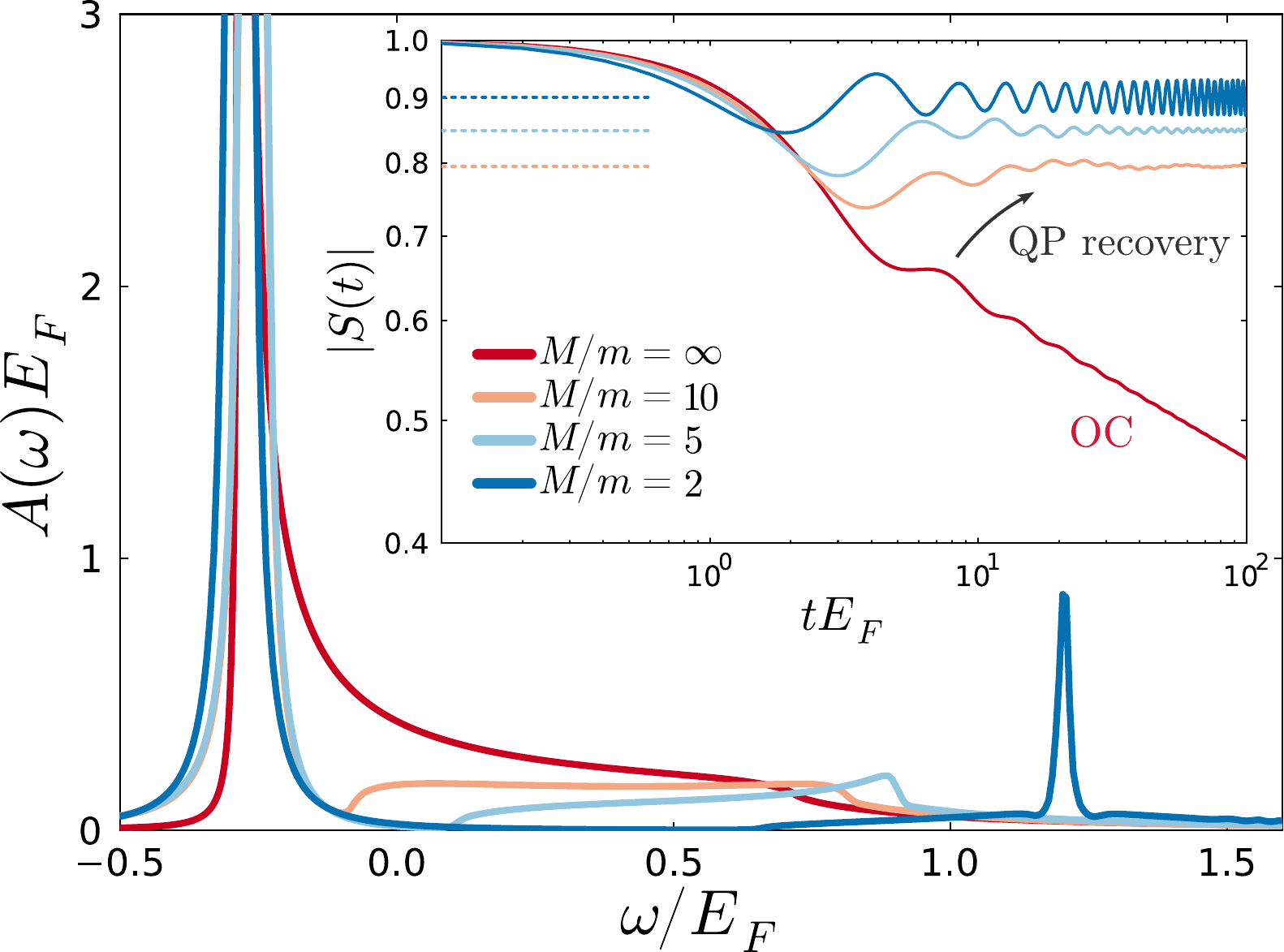}
    \caption{\textbf{Transition from static to mobile impurity.} Spectral function $A(\omega)$ of the Fermi polaron for different mass ratios at inverse scattering length $1/(k_Fa)=-0.5$. Inset: Ramsey signals $|S(t)|$ of each corresponding spectrum, where the dashed lines indicate the quasiparticle weight plateau, $Z = |S(t \to \infty)|^2$.}
    \label{fig:2}
\end{figure}

Particularly noteworthy is the region of strongly suppressed 
spectral weight between the AP peak and the onset of 
the molecule-hole continuum. This \textit{dark 
continuum}, first studied in detail using diagrammatic Monte 
Carlo calculations~\cite{Goulko2016}, also finds a natural 
explanation within the mass-gap picture: the gapped 
dispersion $E_{\mathbf{k}}$ imposes an energy 
cost of $\Delta(M)$ on particle-hole excitations near 
the Fermi surface, kinematically suppressing spectral 
weight in this window. Therefore, the dark continuum provides 
a clear signature of impurity motion --- marking the separation between coherent quasiparticle 
dynamics and incoherent many-body scattering. Its width 
is determined by the separation between the in-gap state and the lower band, and vanishes as 
$M \to \infty$. In this limit, the OC spectrum is recovered with the characteristic power-law behavior and Fermi-edge singularity~\cite{Anderson:1967,Pimenov:2018,Gievers2025}. 

The crossover from quasiparticle to OC physics can be 
followed explicitly in Fig.~\ref{fig:2}, which shows 
$A(\omega)$ for several mass ratios at a fixed negative scattering length, 
$1/(k_F a) = -0.5$. For small $M/m$, the spectrum is 
dominated by a sharp AP peak that carries most of the 
spectral weight. As the mass ratio increases, this peak 
gradually loses weight to the incoherent 
background, while the dark continuum shrinks proportionally to $\Delta(M)$. In the static limit, the 
quasiparticle peak has vanished entirely and is replaced by a power-law Fermi-edge singularity 
characteristic of the OC~\cite{Knap:2012OC}.
The mass-gap FDA thus provides a unified framework 
that smoothly interpolates between these two regimes.
Note that the gapped dispersion shifts the onset of the RP branch towards negative scattering lengths as the mass ratio decreases; as seen by the appearance of the RP resonance at low mass ratios and high frequencies in Fig.~\ref{fig:2}.

%%%%%%%%%%%%%%%%%%%%%%%%%%%%
\setsmalltitle{Ramsey dynamics} In real-time dynamics, the mass-gap serves as a mechanism controlling the 
emergence of a finite quasiparticle weight from the OC. The inset of Fig.~\ref{fig:2} shows the Ramsey signal $S(t)$ 
corresponding to the same mass ratios as for $A(\omega)$, 
displayed on a double-logarithmic scale. In the static limit, 
$S(t)$ exhibits the characteristic power-law decay for $t\to\infty$,
\begin{align}
    S(t) \sim t^{-(\delta_F/\pi)^2} \,,
\end{align}
where $\delta_F \equiv \delta(k_F)$ is the scattering phase 
shift for the static impurity at the Fermi surface \cite{Knap:2012OC}. This power-law decay reflects the generation of the infinitely many gapless particle-hole 
excitations near $k_F$ underlying the OC: each 
excitation costs vanishingly small energy, and their 
cumulative effect destroys the overlap between the 
interacting and non-interacting ground states, leading to ${|S(t \to \infty)|\to 0}$.

For a finite-mass impurity, $\Delta(M)$ imposes a minimum 
energy cost on particle-hole excitations at the Fermi 
surface. As a consequence, $S(t)$ evolves up to a crossover timescale $t^* \sim \Delta(M)^{-1}$, beyond which the signal saturates to a finite plateau. This long-time plateau 
directly encodes the quasiparticle weight, 
${Z = |S(t \to \infty)|^2 > 0}$, (see horizontal dashed lines on the left of the inset in Fig.~\ref{fig:2}), which is the hallmark 
of a well-defined polaron quasiparticle~\cite{Cetina2016}. We note that the behavior 
is analogous to the role of the superconducting gap in 
regularizing the OC for static impurities coupled to 
a BCS 
superfluid~\cite{Wang:2022BCSPRA,Wang:2022BCSPRL,Rodriguez:2025}, 
where the gap likewise cuts off the infrared divergence 
and restores a finite quasiparticle weight.

The Ramsey signal thus provides a direct probe of the crossover from orthogonality catastrophe behavior to coherent quasiparticle formation. At short times $t \ll t^*$, the dynamics 
is dominated by two-body physics, i.e., governed by the phase shift $\delta_F$. At long times 
$t \gg t^*$, the mass gap then asserts itself: the impurity recovers from the OC dynamics and stabilizes as a quasiparticle; reflected in the Ramsey signal as a finite value for $Z$. The value of $Z$ extracted from the plateau is in excellent agreement with the Slater determinant 
calculation of Ref.~\cite{Chen:2025}. Results for a wider range of mass ratios, 
including a comparison to the variational Chevy ansatz~\cite{Chevy2006} 
are provided in the SM~\cite{Supplemental}.

%%%%%%%%%%%%%%%%%%%%%%%%%%%%
\setsmalltitle{Experimental relevance}
Our results are directly relevant for ongoing experiments with either $^{40}$K, $^{41}$K, or $^{133}$Cs impurities in a $^{6}$Li Fermi gas. Here we focus on the latter scenario which corresponds to a mass ratio of $M/m \approx 22$. 
This system represents one of the heaviest realizable Fermi polarons~\cite{Rautenberg:2026}, where finite-mass effects are small but measurable, making it an ideal testbed for distinguishing and benchmarking different theoretical approaches at low temperatures.

\begin{figure}[t]
    \centering
    \includegraphics[width=\linewidth]{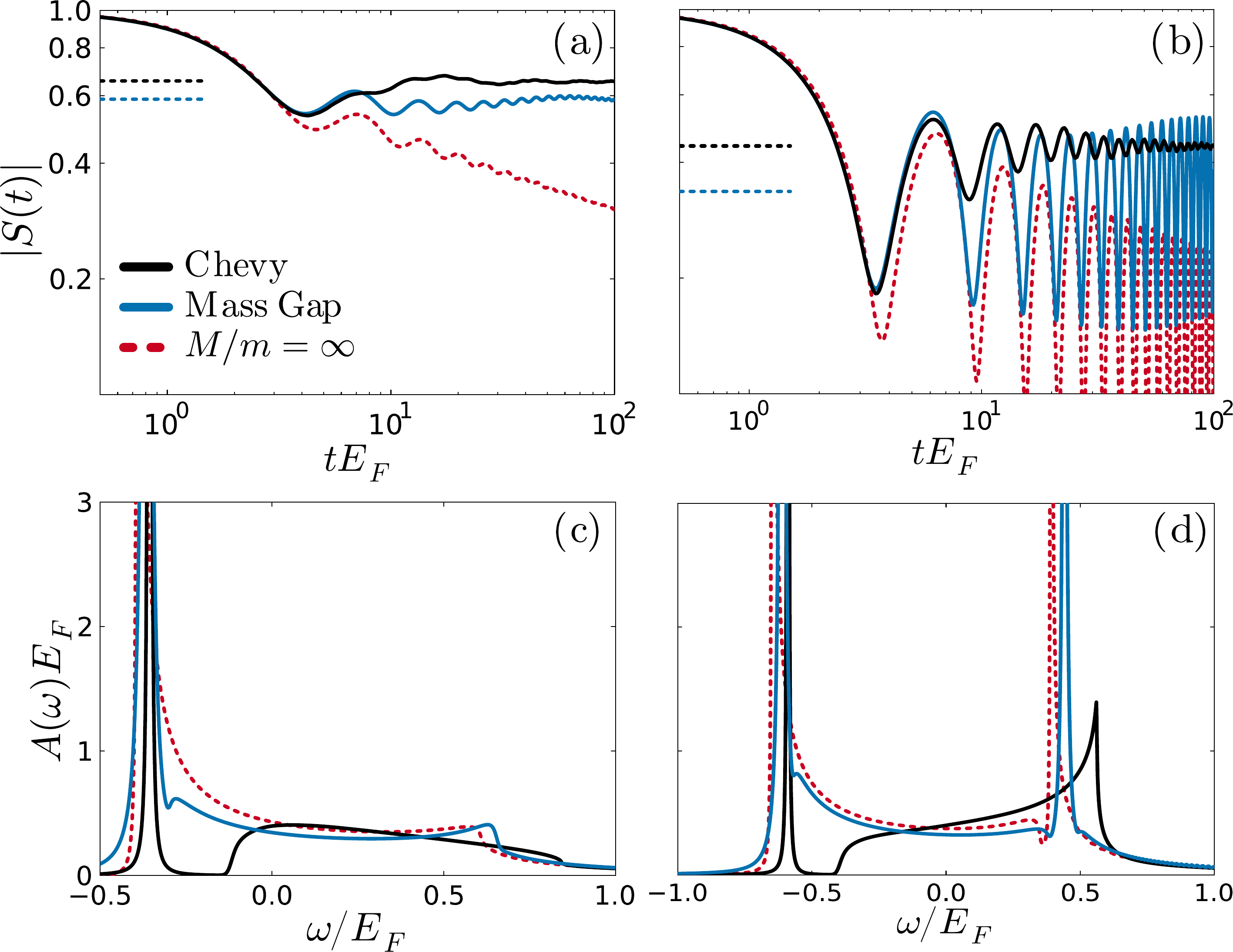}
    \caption{\textbf{Comparison to the variational Chevy ansatz.}
    Ramsey signal $|S(t)|$ for a $^{133}$Cs impurity in a 
    $^{6}$Li Fermi gas ($M/m \approx 22$) at interaction 
    strengths (a) $1/(k_F a) = -0.2$ and (b) 
    $1/(k_F a) = 0.2$, computed from the mass-gap FDA (blue), the Chevy ansatz~\cite{Chevy2006} (black), and the infinite-mass FDA (red). Panels (c) and (d) show the corresponding impurity absorption spectrum $A(\omega)$.}
    \label{fig:3}
\end{figure}

Figures~\ref{fig:3}(a) and (b) show the Ramsey signal $|S(t)|$ computed 
from the mass-gap FDA (blue line) for two representative 
interaction strengths on the attractive 
[$1/(k_F a) = -0.2$, panel (a)] and repulsive 
[$1/(k_F a) = 0.2$, panel (b)] side. For comparison, we also show the corresponding results from the Chevy ansatz~\cite{Chevy2006}, also called truncated basis method~\cite{Parish:2016, Adlong:2020} (black line), and the 
conventional FDA for a static impurity~\cite{Schmidt2018} (red dashed line). The three 
approaches agree well at short times, where the dynamics is governed by the scattering phase shift $\delta_F$. At 
longer times $t > t^*$, however, 
the three theories differ significantly. The static-impurity FDA exhibits a pure power-law decay characteristic of the orthogonality catastrophe, while both the Chevy ansatz and the mass-gap FDA saturate to a finite quasiparticle residue $Z$, consistent with coherent quasiparticle formation at finite mass. However, the mass-gap FDA captures the full OC dynamics before saturating into the long-time plateau. As a result, 
the quasiparticle residue $Z$ lies below the 
Chevy ansatz (horizontal dashed lines), reflecting 
the contribution of infinitely many particle-hole 
excitations. This effect becomes more pronounced 
for heavier impurities, see SM~\cite{Supplemental}.

The plateau in the Ramsey dynamics is a key signature of the mass 
gap in the time domain. The corresponding differences 
in the absorption spectrum $A(\omega)$ are shown in 
Figs.~\ref{fig:3}(c) and (d): while there are only minor differences in the AP peak positions across all three approaches, notable 
differences appear in the repulsive polaron branch, as well as in the separation between the AP peak and the 
incoherent scattering tail. In particular, the Chevy ansatz predicts a stronger suppression of spectral weight that is absent in the mass-gap FDA. Moreover, the RP peak is also suppressed in the Chevy ansatz and shifted to higher energies in comparison to the FDA approaches.

%

%%%%%%%%%%%%%%%%%%%%%%%%%%%%
\setsmalltitle{Conclusion} We have introduced the mass-gap FDA, which extends the conventional FDA to mobile impurities by exploiting the quadratic 
structure of the mass-gap Hamiltonian~\cite{Chen:2025}. 
The method enables exact and computationally efficient 
calculations of impurity spectra and Ramsey 
dynamics within the mass-gap description of Fermi polarons, 
overcoming the fundamental restriction of the FDA to 
static impurities.

Our results explicitly demonstrate a smooth crossover from 
the Fermi-edge singularity of the Anderson orthogonality catastrophe to the 
emergence of a well-defined quasiparticle peak for finite-mass impurities. This crossover is directly 
controlled by the mass gap $\Delta(M)$, which 
regularizes the infrared divergence of the OC and sets 
a characteristic timescale beyond which the Ramsey signal saturates to a finite 
quasiparticle weight $Z$. The mass-gap FDA captures the 
full richness of the impurity spectrum in a unified 
framework, including the dark continuum, the 
molecule-hole scattering continuum, the repulsive polaron, 
and the recovery of quasiparticles in 
the long-time behavior of $S(t)$.

Beyond the results presented here, the mass-gap FDA 
opens the door to a range of systematic extensions. 
Corrections beyond mean-field can be incorporated 
order by order through a controlled perturbative
expansion scheme of the Ramsey signal in powers of 
the quartic LLP terms (see SM~\cite{Supplemental}). The vanishing of the first-order correction demonstrates the power of the mass-gap approach, and
higher-order corrections provide further accuracy towards the mass-balanced 
limit. The framework naturally extends to further non-equilibrium quantum many-body
protocols~\cite{yang2025noneq,Yang:2026_Floquet}, including spin-echo experiments and Rabi-coupled 
polarons~\cite{Knap:2012OC,Adlong:2021,Hu2023,Vivanco2025,LiuZeyu2026Rabi}. Furthermore, the case of finite impurity momentum can be accounted for within the mass-gap FDA, since the corresponding term in the Hamiltonian remains quadratic. The generalization to finite temperature is also possible in principle, but requires further investigation. The mass-gap approach also applies directly 
to lower-dimensional systems such as exciton-polarons 
in two-dimensional 
semiconductors~\cite{Sidler:2017,imamoglu2021exciton}, 
as well as other interaction 
potentials, making it a broadly applicable framework for studying the
quantum dynamics of mobile impurities.

%%%%%%%%%%%%%%%%%%%%%%%%%%%%
\setsmalltitle{Acknowledgements} We thank A.~Christianen and M.~Gievers for helpful discussions and useful comments on the manuscript. We acknowledge funding  by the DFG (German Research Foundation) – Project-ID 273811115 – SFB 1225 ISOQUANT, and under Germany's Excellence Strategy EXC 2181/1 - 390900948 (the Heidelberg STRUCTURES Excellence Cluster). R.S. was supported within the DFG scientific network `A(E)MP - Appearance of the Effective Mass in Polaron Models' (Grant No. 569490025).

\bibliography{ref}
\clearpage
\onecolumngrid

%\appendix
\renewcommand{\theequation}{S\arabic{equation}}
\renewcommand{\thefigure}{S\arabic{figure}}
\begin{center}
\large{\textbf{Supplemental Material for}}\\\large{\textbf{``Mass-gap functional determinant approach for mobile Fermi polarons''}}
\\[1.5ex]
\normalsize{Emilio Ramos Rodr\'iguez*, Eugen Dizer*, Xin Chen, and Richard Schmidt}
\end{center}
In this Supplemental Material, we validate the mass-gap model within a perturbative functional determinant approach (FDA). To this end, starting from the full LLP Hamiltonian, we treat the quartic interaction perturbatively in $m/M$. We first prove Wick's theorem on the FDA contour, which provides the foundation for the subsequent mean-field treatment. We then show that the first-order perturbative correction in $m/M$ vanishes identically, confirming that the mass-gap Hamiltonian is exact to this order. A mean-field analysis of the quartic term yields a time-dependent quadratic Hamiltonian that contributes only to the $l=1$ angular-momentum channel, whose net effect on the FDA signal is of order $\mathcal{O}\big((m/M)^2\big)$. These results  establish the mass-gap model as the leading-order description of the Fermi polaron. In the second part of this Supplemental Material, we provide details on the spectra and Ramsey signals obtained from the variational Chevy ansatz and compare with the mass-gap FDA.

\section{Perturbative FDA}
The focus of this section is to establish a perturbative approach to the FDA for mobile impurities. As mentioned in the main text, in linear response theory, the absorption spectrum $A(\omega)$ can be obtained by evaluating Fourier transform of the Ramsey interference signal, $S(t)=\text{Tr}\left[\hat{\rho}\ \ee^{i\hat{H}_0t}\ee^{-i\hat{H}t}\right]$ with $\hat{\rho}$ the density matrix. The time-resolved signal, $S(t)$, can only be exactly evaluated if the Hamiltonians for the non-interacting ($\hat{H}_0$) and interacting ($\hat{H}$) evolution have a bilinear form. For the Fermi polaron, this is not the case, since $\hat{H}$ has the expression
\begin{equation}\label{eq:ham_int}
    \hat{H}=\frac{\hat{\mathbf{P}}^2}{2M} + \sum_{\mathbf{k}} \epsilon_{\mathbf{k}}\, \hat{c}^{\dagger}_{\mathbf{k}}\hat{c}_{\mathbf{k}} + \frac{g}{\mathcal{V}} \sum_{\mathbf{k},\mathbf{q}} \ee^{i\mathbf{q}\cdot\hat{\mathbf{r}}}\, \hat{c}^{\dagger}_{\mathbf{k}-\mathbf{q}}\hat{c}_{\mathbf{k}}\ ,
\end{equation}
where $\hat{\mathbf{P}}$ and $\hat{\mathbf{r}}$ are the momentum and position operators of the impurity, respectively. As detailed in the Supplemental Material of~\cite{Chen:2025}, the impurity can be decoupled by means of the Lee-Low-Pines (LLP) transformation, $ \hat{U} = \ee^{i\hat{\mathbf{r}}\cdot\hat{\mathbf{P}}_f}$, with $\hat{\mathbf{P}}_f=\sum_{\mathbf{k}}\mathbf{k}\hat{c}^{\dagger}_{\mathbf{k}}\hat{c}_{\mathbf{k}}$. Applying the transformation to Eq.~\eqref{eq:ham_int}, and with the respective operator re-ordering,  yields the LLP Hamiltonian~\cite{Chen:2025}: 
\begin{equation} \label{eq:effective-Hamiltonian-reordered}
    \hat{\cal H}_{\mathrm{LLP}} = \hat{U} \hat{H}\hat{U}^{-1}=\, \, \frac{\mathbf{P}^2}{2M}+\sum_{|\mathbf{q}|<k_F} \frac{\mathbf{q}^2}{2M} + \hat{\cal H}_0(\mathbf{P}) + \frac{g}{\mathcal{V}} \sum_{\mathbf{k},\mathbf{k}'} \hat{c}^{\dagger}_{\mathbf{k}}\hat{c}_{\mathbf{k}'} + \hat{\cal{H}}_{\mathrm{int}}\ ,
    \end{equation}
where 
\begin{equation}\label{eq:H0def}
    \hat{\cal H}_0(\mathbf{P})=\sum_{\mathbf{k}} \left(E_{\mathbf{k}}-\frac{\mathbf{k}\cdot\mathbf{P}}{M}\right)\hat{c}^{\dagger}_{\mathbf{k}}\hat{c}_{\mathbf{k}} 
\end{equation}
and
\begin{equation}\label{eq:quarticHint}
    \hat{\cal{H}}_{\mathrm{int}}= \frac{1}{2M} \sum_{|\mathbf{q}|,|\mathbf{q}'|<k_F} \left(\mathbf{q}\cdot\mathbf{q}'\right) \hat{c}_{\mathbf{q}}\hat{c}_{\mathbf{q}'} \hat{c}^{\dagger}_{\mathbf{q}'}\hat{c}^{\dagger}_{\mathbf{q}} - \frac{1}{M} \sum_{|\mathbf{q}|<k_F,|\mathbf{k}|>k_F} \left(\mathbf{k}\cdot\mathbf{q}\right) \hat{c}^{\dagger}_{\mathbf{k}}\hat{c}_{\mathbf{q}}\hat{c}^{\dagger}_{\mathbf{q}}\hat{c}_{\mathbf{k}} + \frac{1}{2M} \sum_{|\mathbf{k}|,|\mathbf{k}'|>k_F} \left(\mathbf{k}\cdot\mathbf{k}'\right) \hat{c}^{\dagger}_{\mathbf{k}}\hat{c}^{\dagger}_{\mathbf{k}'}\hat{c}_{\mathbf{k}'}\hat{c}_{\mathbf{k}} \,.
\end{equation}
Here, the reordering of creation and annihilation operators is performed in such a way that it follows $\hat{\cal H}_{\mathrm{int}}\ket{\text{FS}}=0$. 

The momentum ${\bf P}$ in \cref{eq:effective-Hamiltonian-reordered} maps to the total momentum operator, $\hat{\bf P} + \hat{\bf P}_f$, in \cref{eq:ham_int} under the inverse LLP transformation. Since the total momentum is a good quantum number, the ${\bf P}$ in \cref{eq:effective-Hamiltonian-reordered} can be replaced by a constant number. Fixing ${\bf P}$ to a constant value imposes a strong constraint on the total Hilbert space: it restricts the dynamics to a single total-momentum sector. This restriction hinders a direct generalization to finite temperature, where the thermal density matrix generally involves a mixture of all momentum sectors. By contrast, the zero-temperature density matrix naturally resides within a single sector of definite ${\bf P}$, making translational symmetry a strong symmetry~\cite{Lessa2025S-W}. In the following, we therefore focus on the zero-temperature limit.

One can apply the same LLP prescription to the non-interacting Hamiltonian $H_0$, obtaining a similar expression,
\begin{equation}
     \hat{U}\hat{H}_{\mathrm{0}}\hat{U}^{-1} =\, \, \frac{\mathbf{P}^2}{2M}+\sum_{|\mathbf{q}|<k_F} \frac{\mathbf{q}^2}{2M} +  \hat{\cal H}_0(\mathbf{P}) + \hat{\cal{H}}_{\mathrm{int}}\ .
\end{equation}
Observe that, in the non-interacting case, it follows $[\hat{U}\hat{H}_0\hat{U}^{-1},\hat{\cal H}_{\mathrm{int}}]=0$. At zero temperature, the density matrix reduces to the projection to the Fermi sea, since the Fermi sea has vanishing total momentum, i.e. it is invariant under the LLP transformation. Applying the LLP transformation to the Ramsey signal, one obtains
\begin{equation}
S(t)=\text{Tr}\left[\hat{U}\hat{\rho}\,\hat{U}^{-1}\hat{U}\ee^{i\hat{ H}_0t}\hat{U}^{-1}\hat{U}\ee^{-i\hat{H}t}\hat{U}^{-1}\right]\stackrel{T=0}{=}\Bigl\langle \text{FS}\Bigl|\ee^{i\hat{\cal H}_0(\mathbf{P})t}\ee^{-i\left[\hat{\cal H}(\mathbf{P})+\hat{\cal H}_{\mathrm{int}}\right]t}\Bigl|\text{FS}\Bigl\rangle\ ,
\end{equation}
where $|\text{FS}\rangle$ is the non-interacting Fermi sea.

The exact FDA \cite{Klich2003} is inherently restricted to quadratic Hamiltonians, such as the mass-gap model considered here. To incorporate higher-order effects, one must include the quartic terms in the field operators of the Fermi gas, as specified in \cref{eq:quarticHint}. In order to treat these quartic interactions perturbatively, we first write the full Ramsey signal as
\begin{align} \label{eq:full-Ramsey-signal}
    S(t) = \Bigl\langle \text{FS}\Bigl | \ee^{i\hat{\mathcal{H}}_0 t} 
    \ee^{-i\hat{\mathcal{H}} t}\, \mathcal{T}\exp\!\left\{
    -i\int_0^{t} \hat{\mathcal{H}}_{\mathrm{int}}(t')\, 
    dt'\right\}\Bigl | \text{FS}\Bigr\rangle \,,
\end{align}
where $\hat{\mathcal{H}}_{\mathrm{int}}(t) = 
\ee^{i\hat{\mathcal{H}}t}\hat{\mathcal{H}}_{\mathrm{int}}
\ee^{-i\hat{\mathcal{H}}t}$ is the interaction-picture representation with respect to the mass-gap Hamiltonian, and $\mathcal{T}$ denotes the time-ordering operator. Expanding the time-ordered exponential in powers of $\hat{\mathcal{H}}_{\mathrm{int}}$ yields the higher-order corrections to $S(t)$. Each term in this expansion involves an expectation value of a time-ordered product of field operators. To decompose such expectation values into products of Green's functions, one requires Wick's theorem. In the next subsection, we therefore prove that Wick's theorem holds, which in turn motivates a diagrammatic expansion for higher-order corrections to $S(t)$.

\subsection{Wick's theorem for the FDA}\label{sec:Wick}
\begin{figure}[t!]
    \centering
    \includegraphics[width=0.6\textwidth]{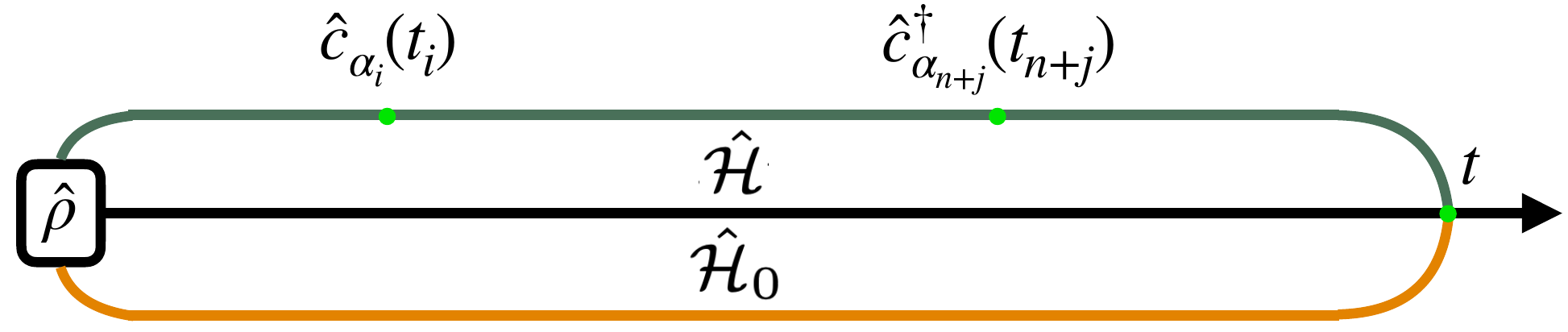}
    \caption{Illustration of the non-unitary FDA contour. The green (orange) contour represents the positive (negative) time evolution under the Hamiltonian $\hat{\mathcal{H}}$ ($\hat{\mathcal{H}}_0$). The operators $\hat{c}_{\alpha_i}(t_i)$ and $\hat{c}^\dagger_{\alpha_{n+j}}(t_{n+j})$ are inserted on the green contour. }
    \label{fig:FDAcontour-illu}
\end{figure}

In this subsection, we provide a sketch of the proof of Wick's theorem within the functional determinant approach (FDA) for arbitrary temperature. Wick's theorem plays a central role in perturbative quantum field theory. While the quadratic action underlying Wick's theorem is well known, explicitly presenting a proof of the theorem is essential for obtaining the overall coefficients for the contractions and the Green's functions. We start with the $2n$-operator correlation function,
\begin{align}
    {\cal C}^{2n}(t_1,\cdots, t_{2n}) := \text{Tr}\left[\hat{\rho}\,\ee^{i\hat{\cal H}_0t}\ee^{-i\hat{\cal H}t}{\cal T} \prod_{i=1}^n\hat{c}_{\alpha_i}(t_i)\hat{c}^\dagger_{\alpha_{n+i}}(t_{n+i})\right],
\end{align}
where ${\cal T}$ denotes the time-ordering operator, and $\hat{c}_{\alpha_i}(t_i)$
$(\hat{c}^\dagger_{\alpha_{n+i}}(t_{n+i}))$ are the annihilation (creation)
operators in the Heisenberg picture with respect to the Hamiltonian
$\hat{\cal H}$. Here, $\alpha_i$ labels the single-particle state of the $i$th
operator, $t_i$ is its time argument, and $t$ is the total evolution time
of the FDA contour [see Fig.~\ref{fig:FDAcontour-illu}]. The density matrix also takes a Gaussian form,
$\hat{\rho}=\exp(-\beta \hat{\cal H}_0)/Z$. Because $\hat{\cal H}_0$ and
$\hat{\cal H}$ are quadratic, the operators $\hat{c}_{\alpha_i}(t_i)$ are
linear combinations of the $\hat{c}_{\alpha_i}(0)$, and the exponential of a
quadratic Hamiltonian can be moved past them. These two properties form the
basis of the proof of Wick's theorem \cite{GAUDIN1960Wick}, given by the expressions:
\begin{subequations}
\begin{align}
    &\hat{c}_{\alpha_i}(t_i):=\ee^{i\hat{\cal H}t_i}\hat{c}_{\alpha_i}\ee^{-i\hat{\cal H}t_i}=\sum_{\beta_i}\left[\ee^{-i\hat{h}t_i}\right]_{\alpha_i\beta_i}\hat{c}_{\beta_i}(0),\\
    &\hat{c}_{\alpha_i}^\dagger(t_i):=\ee^{i\hat{\cal H}t_i}\hat{c}^\dagger_{\alpha_i}\ee^{-i\hat{\cal H}t_i}=\sum_{\beta_i}\left[\ee^{i\hat{h}t_i}\right]_{\beta_i\alpha_i}\hat{c}_{\beta_i}^\dagger(0),\\
    &\hat{\rho}\ee^{i{\hat{\cal H}}_0t}\ee^{-i\hat{\cal H}t}\hat{c}_{\alpha_i}(t_i) = \sum_{\beta_i}\left[\ee^{-i\hat{h}t_i}\ee^{i\hat{h}t}\ee^{-i\hat{h}_0t}\ee^{\beta \hat{h}_0}\ee^{i\hat{h}t_i}\right]_{\alpha_i\beta_i}\hat{c}_{\beta_i}(t_i)\hat{\rho}\ee^{i{\hat{\cal H}}_0t}\ee^{-i\hat{\cal H}t},\\
    &\hat{\rho}\ee^{i{\hat{\cal H}}_0t}\ee^{-i\hat{\cal H}t}\hat{c}^\dagger_{\alpha_i}(t_i) = \sum_{\beta_i}\left[\ee^{-i\hat{h}t_i}\ee^{-\beta \hat{h}_0}\ee^{i\hat{h}_0t}\ee^{-i\hat{h}t}\ee^{i\hat{h}t_i}\right]_{\beta_i\alpha_i}\hat{c}^\dagger_{\beta_i}(t_i)\hat{\rho}\ee^{i{\hat{\cal H}}_0t}\ee^{-i\hat{\cal H}t}\, ,
\end{align}
\end{subequations}
where $\hat{h}_0$ and $\hat{h}$ are the single-particle 
operators of $\hat{\mathcal{H}}_0$ and 
$\hat{\mathcal{H}}$, respectively.

For simplicity, the matrix $\exp(i\hat{h}t)\exp(-i\hat{h}_0t)\exp(\beta \hat{h}_0)$ is denoted as $V$, while $\exp(-i\hat{h}t_i)$ is denoted as $U_i$. In the case where the first operator in the time ordering is an annihilation operator, denoted as $\hat{c}_{\alpha_1}(t_{1})$, the cyclic property of the trace yields an identity for the $2n$-operator correlation function ${\cal C}^{2n}$:
\begin{align}
    C^{2n} = &\left[U_iVU_i^\dagger\right]_{\alpha_1\beta_1}\text{Tr}\left\{\hat{\rho}\,\ee^{i\hat{\cal H}_0t}\ee^{-i\hat{\cal H}t}{\cal T} \hat{c}^\dagger_{\alpha_{n+1}}(t_{n+1})\left[\prod_{i=2}^n\hat{c}_{\alpha_1}(t_i)\hat{c}^\dagger_{\alpha_{n+i}}(t_{n+i})\right]\hat{c}_{\beta_1}(t_1)\right\} \nonumber\\
    =&\left[U_iVU_i^\dagger\right]_{\alpha_1\beta_1}\sum_{k=2}^n\left[\ee^{-i\hat{h}(t_1-t_{n+k})}\right]_{\beta_1\alpha_{n+k}}\text{Tr}\left\{\hat{\rho}\,\ee^{i\hat{\cal H}_0t}\ee^{-i\hat{\cal H}t}{\cal T} \hat{c}^\dagger_{\alpha_{n+1}}(t_{n+1})\hat{c}_{\alpha_k}(t_k)\left[\prod_{i=2,i\neq k}^n\hat{c}_{\alpha_1}(t_i)\hat{c}^\dagger_{\alpha_{n+i}}(t_{n+i})\right]\right\}\nonumber\\
    &+\left[U_iVU_i^\dagger\right]_{\alpha_1\beta_1}\left[\ee^{-i\hat{h}(t_1-t_{n+1})}\right]_{\beta_1\alpha_{n+k}}\text{Tr}\left\{\hat{\rho}\,\ee^{i\hat{\cal H}_0t}\ee^{-i\hat{\cal H}t}{\cal T} \left[\prod_{i=2,i\neq k}^n\hat{c}_{\alpha_1}(t_i)\hat{c}^\dagger_{\alpha_{n+i}}(t_{n+i})\right]\right\}\nonumber\\
    &-\left[U_iVU_i^\dagger\right]_{\alpha_1\beta_1}C^{2n}_{\alpha_1\rightarrow\beta_1},\label{eq:cyclic}
\end{align}
where we have used the anti-commutation relation $\{\hat{c}_{\alpha}(t_\alpha),\hat{c}^\dagger_\beta(t_\beta)\}=[\exp(-i\hat{h}(t_\alpha-t_\beta))]_{\alpha\beta}$. Conversely, if the first operator is a creation operator, we can still apply the cyclicity of the trace to the first annihilation operator by utilizing the anti-commutation relations before permuting it with the exponential of the quadratic Hamiltonian. Notably, \cref{eq:cyclic} relates the $2n$-operator correlators to the $(2n-2)$-operator correlators:
\begin{align}
    {\cal C}^{2n}=&\sum_{i,j}\left\{U_i\Big[(\mathbb{I}+V)^{-1}V \Theta(t_i-t_{n+j})-(\mathbb{I}+V)^{-1}\Theta(t_{n+j}-t_i)  \Big] U_i^\dagger\ee^{-i\hat{h}(t_i-t_{n+j})}  \right\}_{\alpha_i\alpha_{n+j}}{\cal C}_{\hat{i}\hat{j}}^{2n-2}\nonumber\\
    =& \sum_{i,j}\left\{\ee^{-i\hat{h}t_i}\Big[\Theta(t_i-t_{n+j})\mathbb{I}-(\mathbb{I}+V)^{-1}\Big] \ee^{i\hat{h}t_{n+j}}\right\}_{\alpha_i,\alpha_{j+n}} {\cal C}_{\hat{i}\hat{j}}^{2n-2}.\label{eq:deduct}
\end{align}
Following this logic, the $2n$-operator correlator can be decomposed into combinations of contractions. As a result, we obtain the explicit form of the Wick contractions:
\begin{subequations}
    \begin{align}
    &\contraction{ }{\hat{c}_{\alpha}} {(t_\alpha)}{ \hat{c}_{\beta}^\dagger }  \hat{c}_{\alpha}(t_\alpha) \hat{c}_{\beta}^\dagger(t_\beta)=\left\{\ee^{-i\hat{h}t_\alpha}\Big[\Theta(t_\alpha-t_\beta)\mathbb{I}-(\mathbb{I}+\ee^{i\hat{h}t}\ee^{-i\hat{h}_0t}\ee^{\beta \hat{h}_0})^{-1}
    \Big]\ee^{i\hat{h}t_\beta}\right\}_{\alpha\beta},\\
    &{\cal C}^{2n}=\text{Tr}[\hat{\rho}\,\ee^{i\hat{\cal H}_0t}\ee^{-i\hat{\cal H}t}]\sum\{\text{all possible Wick contractions}\}.
\end{align}
\end{subequations}
Specifically, in the zero-temperature limit ($T\rightarrow 0$), the Wick contraction can be simplified to:
\begin{align}
    \contraction{ }{\hat{c}_{\alpha}} {(t_\alpha)}{ \hat{c}_{\beta}^\dagger }  \hat{c}_{\alpha}(t_\alpha) \hat{c}_{\beta}^\dagger(t_\beta)=\left\{\ee^{-i\hat{h}t_\alpha}\Big[ \Theta(t_\alpha-t_\beta)\mathbb{I}- \hat{n}\Big]\ee^{i\hat{h}t_\beta}\right\}_{\alpha\beta},\label{eq:gfFDA}
\end{align}
with the occupation matrix $\hat{n}=\lim_{\beta\rightarrow\infty}(\mathbb{I}+\exp(\beta \hat{h}_0))^{-1}$.

\subsection{First-order perturbation for the mass-gap FDA}
From the Wick contraction in \cref{eq:gfFDA}, the zero-temperature propagator is given by
\begin{align}\label{eq:G_T=0}
    G(t_1,t_2)_{\mathbf{kq}}:= -i \frac{\langle\ee^{i\hat{\mathcal{H}}_0 t} 
    \ee^{-i\hat{\mathcal{H}} t}\, \hat{\mathcal{T}} \hat{c}_{\bf k}(t_1)\hat{c}^\dagger_{\bf q}(t_2)\rangle}{\langle\ee^{i\hat{\mathcal{H}}_0 t} 
    \ee^{-i\hat{\mathcal{H}} t}\rangle} =-i\langle {\bf k}|\ee^{-it_1\hat{h}}[\Theta(t_1-t_2)\mathbb{I}-\hat{n}]\ee^{it_2\hat{h}}|{\bf q}\rangle\ ,
\end{align}
where $|{\bf k}\rangle$ and $|{\bf q}\rangle$ denote single-particle plane-wave states.  For simplicity, we restrict to total momentum ${\bf P}=0$, for which the occupation matrix is $\langle {\bf q}|\hat{n}|{\bf q}'\rangle=\delta_{\mathbf{q},\mathbf{q}'}\Theta(k_F-|{\bf q}|)$. The contact interaction in \cref{eq:ham_int} combined with rotational symmetry constrains the propagator to the form,
\begin{align}\label{eq:GkqAnsatz}
    G(t_1,t_2)_{\mathbf{kq}}=\delta_{{\bf k},{\bf q}}[\Theta(t_1-t_2)-\Theta(k_F-|{\bf k}|)]\ee^{-i\epsilon_{\bf k}(t_1-t_2)} + F(t_1, t_2, |{\bf k}|,|{\bf q}|)\ ,
\end{align}
where $F$ depends only on the magnitudes of ${\bf k}$ and ${\bf q}$. Using the expression for the full Ramsey signal in \cref{eq:full-Ramsey-signal}, the expansion up to first order in $\hat{\mathcal{H}}_{\mathrm{int}}$ takes the form:
\begin{align} \label{eq:Ramsey-signal_firstorder}
    S(t) = \Bigl\langle \text{FS}\Bigl | \ee^{i\hat{\mathcal{H}}_0 t} 
    \ee^{-i\hat{\mathcal{H}} t}\, \Bigl | \text{FS}\Bigr\rangle-i\int_0^{t}\Bigl\langle \text{FS}\Bigl | \ee^{i\hat{\mathcal{H}}_0 t} 
    \ee^{-i\hat{\mathcal{H}} t} \hat{\mathcal{H}}_{\mathrm{int}}(t')\, 
    \Bigl | \text{FS}\Bigr\rangle dt'+\mathcal{O}(\hat{\mathcal{H}}_{\mathrm{int}}^2) \,.
\end{align}
The expectation value of the second term in~\cref{eq:Ramsey-signal_firstorder} decomposes into three independent contributions, see \cref{eq:quarticHint}. Consider first the term with $|{\bf q}|,|{\bf q}'|<k_F$. Applying Wick's theorem yields
\begin{align}
&\langle\ee^{i\hat{\mathcal{H}}_0 t}
\ee^{-i\hat{\mathcal{H}} t} \frac{1}{2M} \sum_{|\mathbf{q}|,|\mathbf{q}'|<k_F} \left(\mathbf{q}\cdot\mathbf{q}'\right) \hat{c}_{\mathbf{q}}(t')\hat{c}_{\mathbf{q}'} (t')\hat{c}^{\dagger}_{\mathbf{q}'}(t')\hat{c}^{\dagger}_{\mathbf{q}}(t')\rangle \nonumber\\=& -\langle\ee^{i\hat{\mathcal{H}}_0 t}
\ee^{-i\hat{\mathcal{H}} t}\rangle \frac{1}{2M} \sum_{|\mathbf{q}|,|\mathbf{q}'|<k_F}({\bf q}\cdot {\bf q}')\left[G(t'+0,t')_{{\bf q}{\bf q}}G(t'+0,t')_{{\bf q}'{\bf q}'}-G(t'+0,t')_{{\bf q}'{\bf q}}G(t'+0,t')_{{\bf q}{\bf q}'}\right]\ .\label{eq:Hintqq'Expec}
\end{align}
Integrating over the solid angles $\Omega_{\bf q}$ and $\Omega_{{\bf q}'}$ eliminates the $G_{{\bf q}{\bf q}}G_{{\bf q}'{\bf q}'}$ term, since ${\bf q}\cdot{\bf q}'$ is odd under directional averaging while the diagonal propagators are isotropic. The remaining contribution reduces to
\begin{align}
\langle\ee^{i\hat{\mathcal{H}}_0 t}
\ee^{-i\hat{\mathcal{H}} t}\rangle \frac{1}{2M}\sum_{|{\bf q}|<k_F} {\bf q}^2\big[1-\Theta(k_F-|{\bf q}|)\big]\big[1+2F(t'+0,t',|{\bf q}|,|{\bf q}|)\big]=0\ ,
\end{align}
which vanishes identically because $1-\Theta(k_F-|{\bf q}|)=0$ for all $|{\bf q}|<k_F$. A similar argument shows that the other two terms in \cref{eq:quarticHint} also vanish. Hence, the first-order perturbative correction from \cref{eq:Ramsey-signal_firstorder} vanishes.

\subsection{Effective mean-field LLP potential}
To evaluate \cref{eq:quarticHint} in the mean-field approximation, we first apply Wick's theorem to decompose the quartic term into contracted contributions and residual fluctuations. For brevity, we introduce the symbol ${\cal R}$ to denote the re-ordering of operators specified in \cref{eq:quarticHint}. The term with $|{\bf q}|,|{\bf q}'|<k_F$ then decomposes as
\begin{align}
    &\frac{1}{2M}{\cal R} \sum_{|\mathbf{q}|,|\mathbf{q}'|<k_F} \left(\mathbf{q}\cdot\mathbf{q}'\right) \hat{c}_{\mathbf{q}}\hat{c}_{\mathbf{q}'} \hat{c}^{\dagger}_{\mathbf{q}'}\hat{c}^{\dagger}_{\mathbf{q}} (t)\nonumber \\
    =& -\frac{1}{M}{\cal R}\sum_{|\mathbf{q}|,|\mathbf{q}'|<k_F} \left(\mathbf{q}\cdot\mathbf{q}'\right)\hat{c}_{\mathbf{q}} \hat{c}^\dagger_{\mathbf{q}'}(t)F(t+0,t,|{\bf q}|',|{\bf q}|)\nonumber\\& - \frac{1}{2M}{\cal R} \sum_{|\mathbf{q}|,|\mathbf{q}'|<k_F} \left(\mathbf{q}\cdot\mathbf{q}'\right) \left[\hat{c}_{\mathbf{q}} \hat{c}^{\dagger}_{\mathbf{q}'}(t)-\contraction{}{\hat{c}_{\bf q}}{}{ \hat{c}_{{\bf q}'}^\dagger}\hat{c}_{\bf q}\hat{c}^\dagger_{{\bf q}'}\right]\left[\hat{c}_{\mathbf{q}'} \hat{c}^{\dagger}_{\mathbf{q}}(t)-\contraction{}{\hat{c}_{{\bf q}'}} {}{ \hat{c}_{{\bf q}}^\dagger } \hat{c}_{{\bf q}'}\hat{c}_{{\bf q}}^\dagger\right]\ .
\end{align}
The mean-field approximation amounts to discarding the fluctuation term. Applying the same procedure to all three sectors of \cref{eq:quarticHint} yields the mean-field Hamiltonian
\begin{align}
    \mathrm{M.F.}[\hat{\cal H}_\mathrm{int}]=&-\frac{1}{M}{\cal R}\sum_{|\mathbf{q}|,|\mathbf{q}'|<k_F} \left(\mathbf{q}\cdot\mathbf{q}'\right)\hat{c}_{\mathbf{q}} \hat{c}^\dagger_{\mathbf{q}'}(t)F(t+0,t,|{\bf q}|',|{\bf q}|) + \frac{1}{M}{\cal R}\sum_{|\mathbf{k}|,|\mathbf{k}'|>k_F} \left(\mathbf{k}\cdot\mathbf{k}'\right)\hat{c}^\dagger_{\mathbf{k}'} \hat{c}_{\mathbf{k}}(t)F(t-0,t,|{\bf k}|',|{\bf k}|)\nonumber\\
    &+\frac{1}{M}\left[{\cal R}\sum_{|{\bf q}|<k_F,|{\bf k}|>k_F}({\bf k}\cdot {\bf q}) c_{\bf k}^\dagger c_{\bf q} F(t-0,t,|{\bf k}|,|{\bf q}|)+\mathrm{h.c.}\right] \,.
\end{align}
From \cref{eq:GkqAnsatz}, one obtains the symmetry relations
\begin{subequations}
    \begin{align}
        &F(t+0, t, |{\bf q}|',|{\bf q}|) = F(t+0, t, |{\bf q}|,|{\bf q}|')^* \,, \\
        &F(t-0,t,|{\bf k}|',|{\bf k}|)=F(t-0,t,|{\bf k}|,|{\bf k}|')^* \,,\\
        &F(t-0,t,|{\bf k}|,|{\bf q}|)=F(t-0,t,|{\bf q}|,|{\bf k}|)^* \,,
    \end{align}
\end{subequations}
which shows that $\mathrm{M.F.}[\hat{\cal H}_\mathrm{int}]$ is hermitian. The resulting effective mean-field LLP Hamiltonian is a time-dependent single-particle potential that acts exclusively on the $l=1$ angular-momentum sector. As a consistency check, $\langle \ee^{i\hat{\mathcal{H}}_0 t} \ee^{-i\hat{\mathcal{H}} t}\,\mathrm{M.F.}[\hat{\cal H}\mathrm{int}]\rangle=0$, in agreement with the conclusion of the previous subsection. This implies that the next non-vanishing contribution from the mean-field LLP potential is ${\cal O}[(m/M)^2]$.

\section{Comparison to the variational Chevy ansatz}
In this section, we present additional results for large mass ratios, where the mass-gap FDA becomes exact to leading order in $m/M$, and compare with the variational Chevy ansatz~\cite{Chevy2006}. The Chevy ansatz is based on a variational wave function that truncates the impurity wave function at a single particle-hole excitation of the Fermi sea. One can show that the Chevy ansatz is equivalent to the non-selfconsistent $T$-matrix approach~\cite{Combescot2007}. At zero temperature and zero total momentum, the $T$-matrix is given by
\begin{align} \label{eq:Tmatrix}
    T^{-1}(\mathbf{q},\omega) = \frac{m_r}{2\pi a}
    - \frac{1}{\mathcal{V}} \sum_{|\mathbf{k}|>k_F}\frac{1}{\omega-\epsilon_{\mathbf{k}}-\epsilon^{I}_{\mathbf{q-k}}+i0^{+}} - \frac{1}{\mathcal{V}} \sum_{\mathbf{k}} \frac{2m_r}{k^2} \,,
\end{align}
with reduced mass $m_r = mM/(m+M)$. Here, $\epsilon_{\mathbf{k}}=\mathbf{k}^2/(2m)$ and $\epsilon^I_{\mathbf{k}}=\mathbf{k}^2/(2M)$ are the bath fermion and impurity dispersion relations, respectively. The impurity self-energy at zero momentum is given by
\begin{align} \label{eq:selfenergy}
    \Sigma(\omega) = \frac{1}{\mathcal{V}} \sum_{|\mathbf{q}|<k_F} T(\mathbf{q},\,\omega+\epsilon_{\mathbf{q}}) \,.
\end{align}
The Ramsey signal $S(t)$ is the Fourier transform of the impurity absorption spectrum,
\begin{align} \label{eq:spectral-function}
    A(\omega) = -\frac{1}{\pi}\,\mathrm{Im}\,\frac{1}{\omega-\Sigma(\omega)+i0^{+}} \,.
\end{align}
In practice, a finite broadening $0^{+} \rightarrow 0.0015\,E_F$ is used for all spectra to resolve the quasiparticle peaks.

\begin{figure}[htbp]
    \centering
    \includegraphics[width=0.48\textwidth]{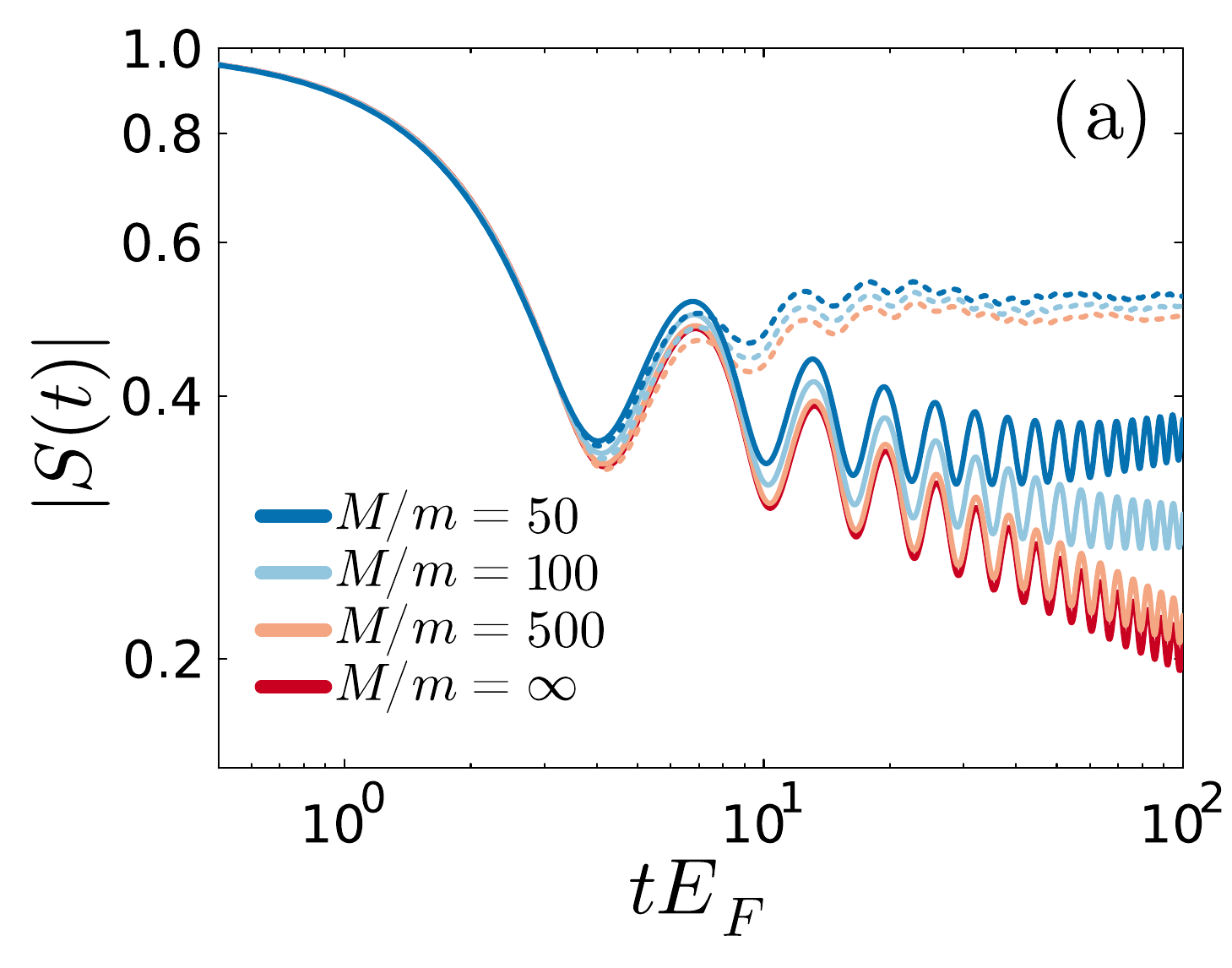}\includegraphics[width=0.48\textwidth]{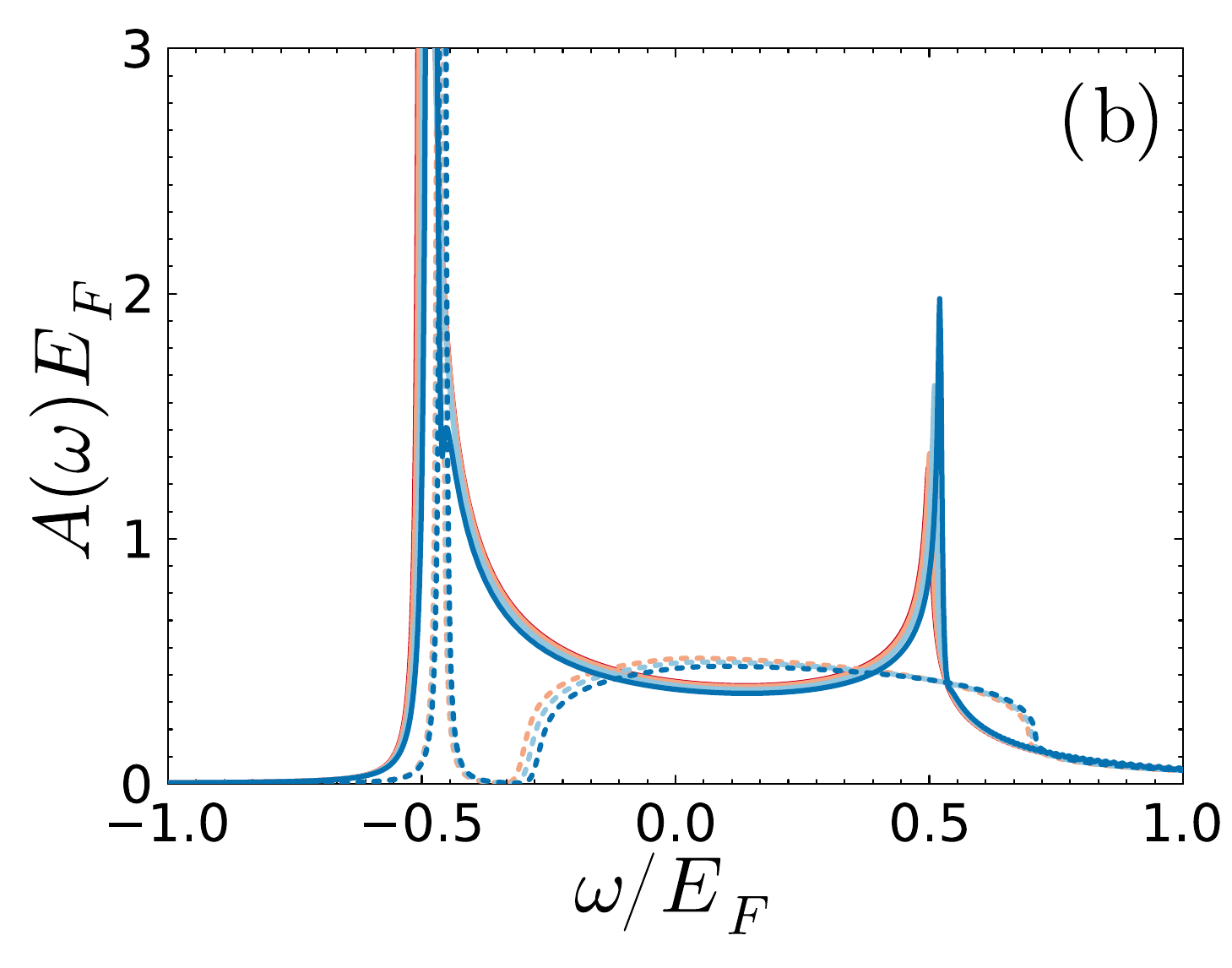}
    \caption{Numerical results for (a) the Ramsey signal $S(t)$ and (b) absorption spectrum $A(\omega)$ obtained from the Chevy ansatz (dashed lines) and the mass-gap FDA (solid lines) at unitarity, $1/(k_Fa)=0$, approaching the heavy impurity limit.}
    \label{fig:Chevy-comparison}
\end{figure}

\begin{figure}[htbp]
    \centering
    \includegraphics[width=0.93\textwidth]{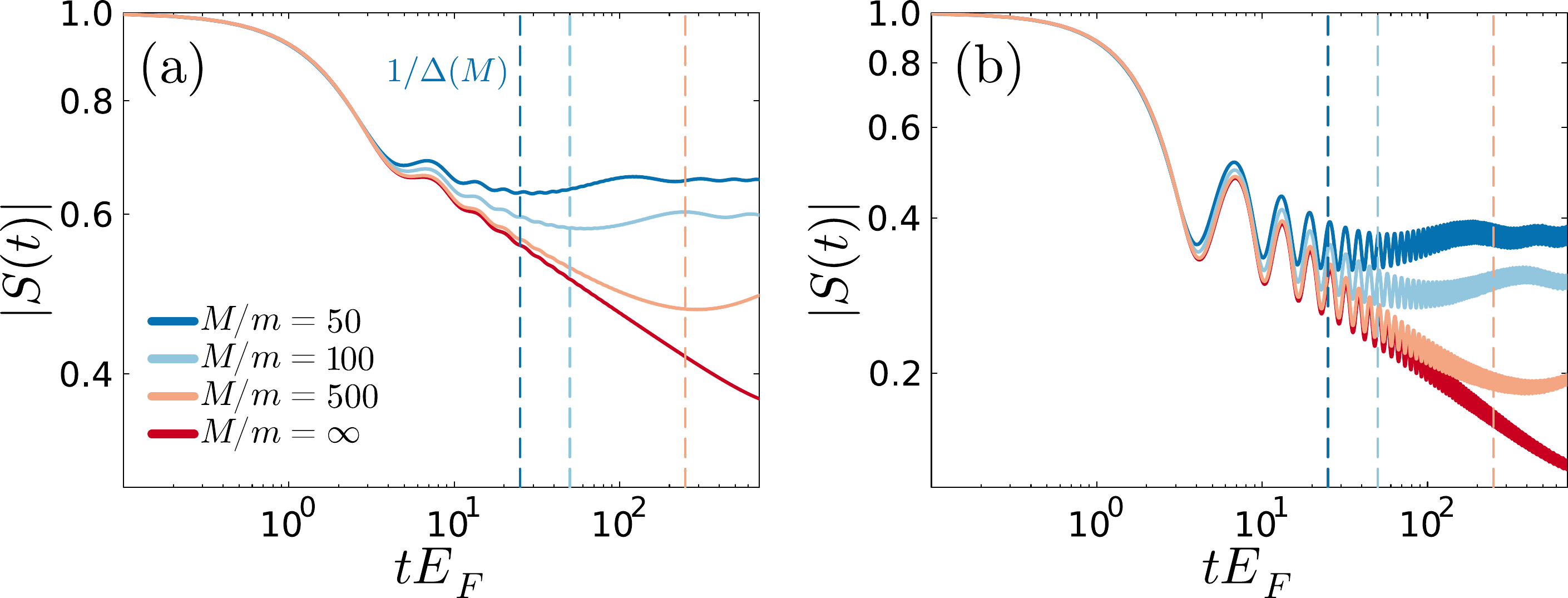}    \caption{Numerical results for the Ramsey signal $S(t)$ at inverse scattering lengths of (a) $1/(k_Fa)=-0.5$ and (b) $1/(k_Fa)=0$ for larger times. Vertical dashed lines mark the characteristic time $1/\Delta(M)$.}
    \label{fig:ramsey_heavy}
\end{figure}

In Fig.~\ref{fig:Chevy-comparison}, we compare the results from the Chevy ansatz (dashed lines) and the mass-gap FDA (solid lines) at unitarity, $1/(k_Fa)=0$, approaching the heavy impurity limit. Figure~\ref{fig:Chevy-comparison}(a) shows the corresponding Ramsey signals. Here, the Chevy ansatz fails to capture the power-law decay in the long-time regime, and the signals tend to a similar quasiparticle plateau. In stark contrast, the mass-gap FDA captures the OC dynamics for times $t<1/\Delta(M)$, ultimately saturating into the long-time quasiparticle plateau. For increasing mass ratios, this plateau gradually decreases into the power-law decay of the infinite-mass limit (red curve), see also Fig.~\ref{fig:ramsey_heavy}. The mass-gap FDA is able to capture the power-law decay of the quasiparticle weight, $Z\propto(M/m)^{-\alpha}$~\cite{Chen:2025}. 

The differences between the two approaches are also evident in the corresponding spectra [see Fig.~\ref{fig:Chevy-comparison}(b)]. The Chevy ansatz exhibits a clear separation between the AP branch and the scattering continuum, without a clear presence of the RP resonance. In the mass-gap FDA spectrum, however, the AP and MHC merge to form the power-law tail of the Fermi-edge singularity. Moreover, the RP resonance is present and carries a higher spectral weight.

%\bibliography{ref}

\end{document}